\documentclass[twocolumn,prl,superscriptaddress,showpacs,preprintnumbers,amsmath,amssymb]{revtex4-1}
\usepackage{graphicx}
\usepackage{color}

\begin{document}

\title{Two-dimensional melting under quenched disorder}

\author{Sven Deutschl\"ander}
\affiliation{Fachbereich f\"ur Physik, Universit\"at Konstanz, D-78464 Konstanz, Germany}
\author{Tobias Kruppa}
\affiliation{Institut f\"ur Theoretische Physik II: Heinrich-Heine-Universit\"at D\"usseldorf,
D-40225 D\"usseldorf, Germany}
\author{Hartmut L\"owen}
\affiliation{Institut f\"ur Theoretische Physik II: Heinrich-Heine-Universit\"at D\"usseldorf,
D-40225 D\"usseldorf, Germany}
\author{Georg Maret}
\affiliation{Fachbereich f\"ur Physik, Universit\"at Konstanz, D-78464 Konstanz, Germany}
\author{Peter Keim}
\affiliation{Fachbereich f\"ur Physik, Universit\"at Konstanz, D-78464 Konstanz, Germany}
\date{\today}

\begin{abstract}

We study the influence of quenched disorder on the two-dimensional melting behavior by using both video-microscopy of superparamagnetic colloidal particles and computer simulations of repulsive parallel dipoles. Quenched disorder is provided by pinning a fraction of the particles. We confirm the occurrence of the Kosterlitz-Thouless-Halperin-Nelson-Young scenario with an intermediate hexatic phase. While the fluid-hexatic transition remains largely unaffected by disorder, the hexatic-solid transition shifts towards lower temperatures for increasing disorder resulting in a significantly broadened stability range of the hexatic phase. In addition, we observe spatio-temporal critical(-like) fluctuations consistent with the continuous character of the phase transitions.

\end{abstract}

\pacs{82.70.Dd, 64.70.D-, 61.20.Ja, 64.70.pv}	

\maketitle

%
%
Since the seminal work of Kosterlitz, Thouless \cite{kosterlitz1972,kosterlitz1973},
Halperin, Nelson and Young  (KTHNY) \cite{Halperin1978,nelson1979,young1979} it is
known that melting in two spatial dimensions can be qualitatively different from three-dimensional
bulk melting. While the latter is typically a phase transition of first order, a two-stage
scenario with an intervening hexatic phase can emerge in two-dimensional
systems which is separated from the fluid and solid phase by two continuous transitions
\cite{strandburg1988}. The KTHNY melting scenario further predicts that in two dimensions,
the melting process is mediated by the unbinding of thermally activated topological defects. In particular,
the emergence of the hexatic phase is related to the dissociation of dislocation pairs into isolated dislocations
\cite{kosterlitz1973,Chaikin_Lubensky}. These break translational symmetry, leading to a vanishing shear modulus. However, the \textit{orientational} symmetry remains quasi-long-range and a finite rotational stiffness is provided.
It has been shown that the KTHNY scenario is realized for soft long-range pairwise potentials  scaling with the inverse cube of the particle separation \cite{lin2006, gribova2011}. In fact, video microscopy experiments with
superparamagnetic colloidal particles pending at a two-dimensional air-water interface and exposed to an external magnetic field perpendicular to the interface have confirmed the KTHNY scenario in detail
\cite{Zahn2000, gruenberg2004,keim2007}. But in systems with a short-range particle interaction, first-order characteristics were found for both transitions \cite{rice1996}.

Typically, two-dimensional melting does not occur under pure bulk conditions but is influenced by quenched (i.e. frozen-in) disorder. Crystallization usually occurs on substrates (examples include graphene sheets, see \cite{meyer2007}) which introduce quenched disorder due to some roughness. The same holds for flux lines pinned by impurities \cite{larkin1970,imry1975,larkin1979,fisher1991} which leads to large critical fields in type II superconductors. Defects may also affect the phase behavior of freely suspended liquid crystal films \cite{geer1992}, of synthetic \cite{vis1995} and biological \cite{pet1990} Langmuir Blodgett films or even 2D protein crystals \cite{berge1991}. Based on a topological defect analysis for weak disorder, Nelson and coworkers \cite{Nelson1983,sachdev1984} have  predicted that the  KTHNY scenario persists with a widening of the hexatic stability range for increasing strength of quenched disorder. This notion was questioned
in subsequent theoretical studies \cite{serota1986}. More recent experimental efforts
\cite{kusner1994,pertsinidis2008,yunker2010,hartmann2010}, simulations  \cite{cha1995,herreravelarde2009,kawasaki2011} and theories \cite{carpentier1998}
have markedly increased our understanding of two-dimensional melting under disorder, but the occurrence
of the hexatic phase was never resolved in all of these studies. Therefore, the above-mentioned
predictions of Nelson  and coworkers \cite{Nelson1983,sachdev1984} have
never been tested by experiment or simulations.

In this Letter, we propose an experiment on superparamagnetic colloids on a glass substrate on which a small fraction of the particles is pinned, inducing quenched disorder.
Clearly, as a reference, the KTHNY scenario occurs for the pure case without any disorder \cite{Zahn2000,keim2007}. We can now systematically study the melting scenario in detail for different fractions of pinned particles.
In our experiments, we confirm the KTHNY scenario and the predictions by
Nelson  and coworkers \cite{Nelson1983,sachdev1984} under disorder. The stability range of the hexatic phase widens upon increasing disorder as opposed
to the prediction of ref.\ \cite{serota1986}. We also perform two-dimensional computer
simulations for parallel dipoles and find good agreement with our experimental data.
Extracting an ``effective'' $K_A$, we recover the scaling of the elasticity modulus in the presence of disorder. Thereby, we provide evidence that melting in the presence of disorder is governed by the same defect-mediated process predicted and confirmed for pure systems. Furthermore, we observe heterogeneous orientational order close to the melting temperature but a long time analysis reveals that such heterogeneities fluctuate strongly on timescales larger than the orientational correlation time indicating critical behavior.

%
%
The experimental system consists of superparamagnetic colloidal particles which are confined in two dimensions and subject to quenched disorder embodied by a random distribution of fixed particles. The colloidal suspension is kept at room temperature and an external magnetic field $H$ applied perpendicular to the particle layer induces a repulsive dipole-dipole potential. The phase behavior is studied by tuning the interaction strength via the external magnetic field, quantified by the dimensionless interaction parameter
\begin{equation}\label{gamma}
 \Gamma=\frac{\mu_0\left(\pi n\right)^{3/2}\left(\chi H\right)^2}{k_BT}\ ,
\end{equation}
with $n$ the 2D particle density, $\chi$ the magnetic susceptibility \cite{chi} and $k_BT$ the thermal energy. The particles have a diameter $d=4.5\:\mu\textrm{m}$ and the mass density $1.7\:\textrm{kg/dm}^3$. The suspension is sealed within a cell, consisting of two parallel cover slips glued together via a hollow cylindrical glass spacer of 5 mm diameter. By gravity, the particles sediment and form a monolayer on the bottom glass plate, where a short-time lateral diffusion constant of $D=0.0295\:\mu\textrm{m}^2/$s is observed. Due to van-der-Waals interactions and chemical reactions between colloids and the glass surface, a small amount of particles pin to the substrate. This distribution is slowly altered by thermal tearing or the creation of new pinning connections, but the pinned particles are fixed on the timescale of our measurements. We exemplify three different sample regions with varying pinning strengths ranging from approx. 0.5 \% to 0.8 \%. The colloidal ensemble is melted from an equilibrated crystalline state by decreasing $H$ in small steps. After each step, the system is allowed to equilibrate for at least 24 hours before particle trajectories are recorded via video microscopy \cite{Ebert2009} for 2.7 hours which equals $\approx50\:\tau_B$.

%
%
Complementarily, computer simulations are carried out, at which the total particle number is fixed to $N = 16000$ and periodic boundary conditions are applied. Each pinning strength is sampled with at least 15 statistically independent configurations of obstacles, which are achieved by pinning randomly selected particles in a fluid configuration of hard disks at a packing fraction of 0.25\%. Within statistical precision, this realization of pinning corresponds to the distribution of pinned particles observed in the experiment. Using the standard Metropolis Monte Carlo algorithm, a full freezing and melting cycle is conducted for each particular setup at which the initially chosen particles remain pinned. After incrementing $\Gamma \propto 1/T$, the system is equilibrated for $5\cdot10^5$ MC sweeps before recording data. While Monte Carlo methods are known to converge rapidly towards static equilibrium states, the underlying phase-space sampling provides a suitable means to study dynamic properties, as well \cite{binder1997}.
For each parameter set of temperature and pinning strength, the observables obtained by MC simulations are averaged over
all sample realizations of disorder.

%
%
The KTHNY theory predicts a two-step melting process, in which isotropic fluid and solid phase are separated by an intermediate hexatic phase. While translational order is only short-range in the hexatic phase, orientational order persists. More precisely, it switches from long-range in the solid over quasi-long-range in the hexatic phase to short-range in the isotropic fluid. The characteristic range of orientational order in the different phases can be quantified in terms of the correlation $g_6(r,t)$ of the bond order parameter
\begin{equation}\label{psi6}
	\psi_{6}=\frac{1}{n_j}\sum_{k}e^{i6\theta_{jk}}\ ,
\end{equation}
where the sum goes over all $n_j$ nearest neighbors of particle $j$, and $\theta_{jk}$ is the angle of the $k$th bond in respect to a certain reference axis. Mapping the characteristic ranges of the \textit{spatial} orientational order on the time domain, we can study the \textit{dynamical} orientational correlation
$g_6(t)=\langle \psi_{6}^*(t)\psi_{6}(0)\rangle$
which, analogous to the spatial correlation, decays exponentially in the isotropic fluid, algebraically in the hexatic phase and approaches a constant value in the solid \cite{nelson_book1983}. This quantity is well suited to characterize the melting process, as discussed in \cite{lin2006,gribova2011} and successfully employed experimentally in \cite{Zahn2000}.
%
%

\begin{figure}[t]
\centering
\includegraphics[width=.95\linewidth]{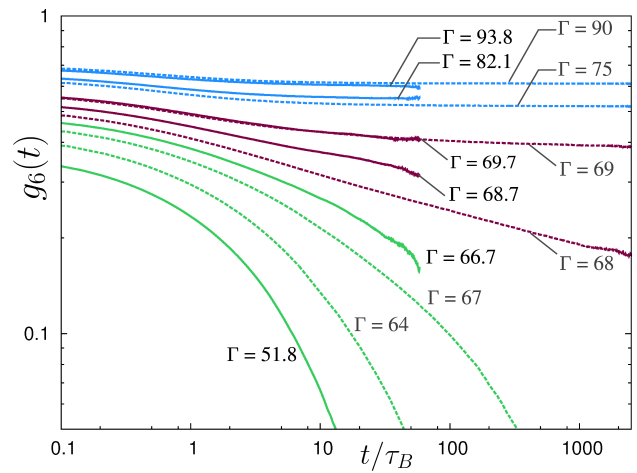}
\caption{Temporal bond orientational correlation function $g_6(t)$ in the presence of quenched disorder plotted versus reduced time $t/\tau_B$ on a double-logarithmic scale.
The fraction of pinned particles is 0.48\% in the experiment and 0.5\% in the simulation. Exemplary curves are shown for the isotropic fluid (green), hexatic (red), and solid (blue) phase, where experimental data is drawn with solid, computer simulations with dashed lines.}
\label{fig_g6t}
\end{figure}
%
%
Fig. \ref{fig_g6t} shows $g_6(t)$ for both, experiment (0.48\% pinning) and simulation (0.5\%). The time axis is reduced to the Brownian time scale $\tau_B=(d/2)^2/D$. After a short-time decay due to Brownian motion, the characteristic behavior of the solid, the hexatic (linear decay in the log-log plot) and isotropic fluid is clearly distinguishable at long times. To confirm the characteristic decay behavior, $g_6(t)$ is fitted with a second order polynomial fit on a double-logarithmic scale:
$\text{ln}(g_6(t)) = a + b\ \text{ln}(t/\tau_B) + c\ \text{ln}^2(t/\tau_B)$,
with dimensionless coefficients $a$, $b$ and $c$. Solid, hexatic and isotropic fluid phases are characterized by the relative contribution of positive or negative curvature, expressed by $c/|b|$. We define an upper and a lower threshold value for $c/|b|$ to distinguish between the negatively curved exponential decay of $g_6(t)$ in the isotropic fluid, positive curvature in the solid, and a linear course in between, reflecting the hexatic phase (for further details see the supplemental material).
%
%
\begin{figure}[b]
\centering
\includegraphics[width=.95\linewidth]{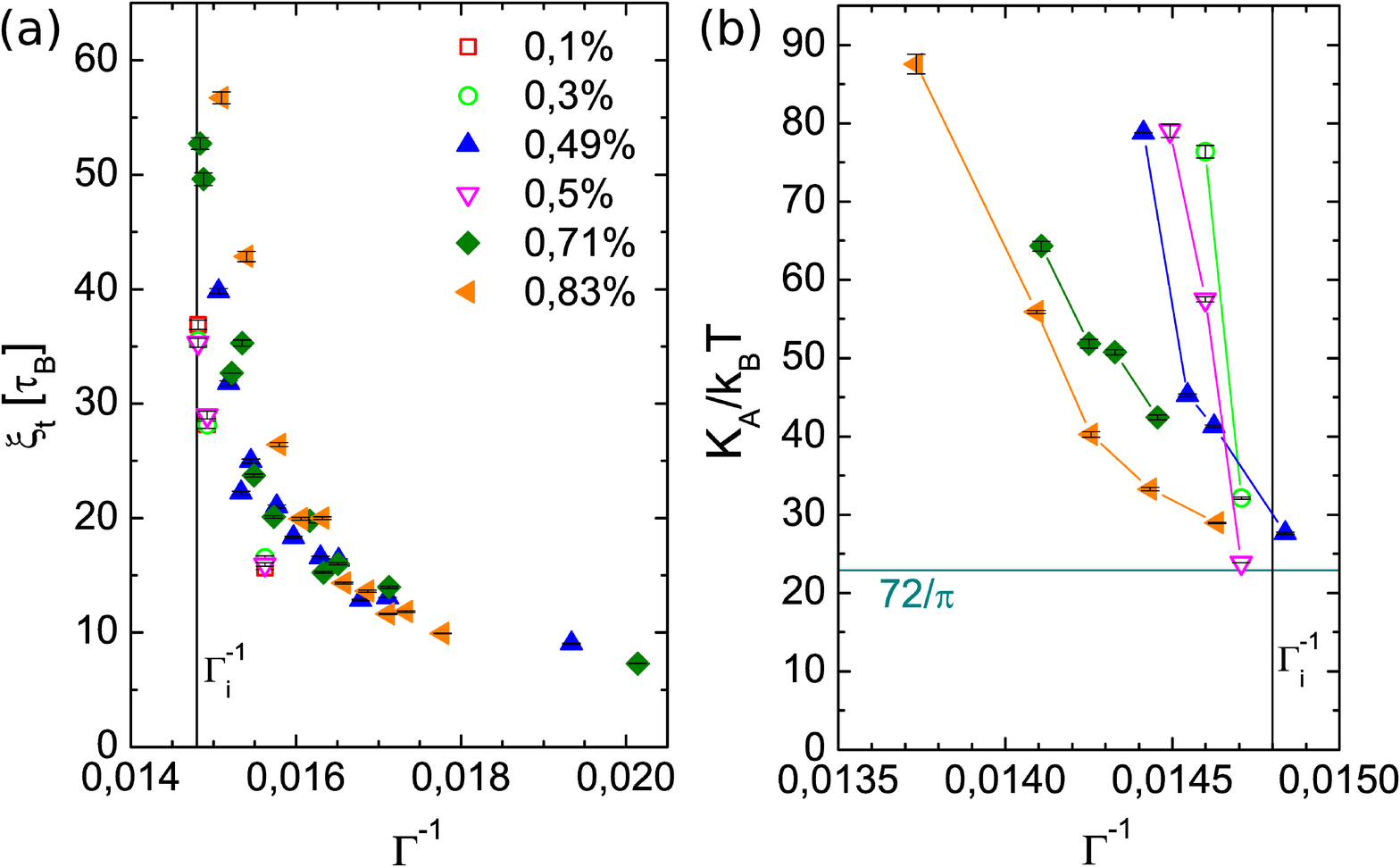}\\
\caption{(a): Orientational correlation time $\xi_t$ and (b): Frank's constant $K_A$, for different concentrations of pinned particles. Filled symbols represent experimental data, open symbols simulation. The meaning of the symbols is the same in (a) and (b), lines are guides to the eye. While $\xi_t$ is almost not affected by different pinning strengths, $K_A$ is clearly lowered with increasing pinning.}
\label{fig_xi}
\end{figure}

To illustrate the critical behavior at the transition points, we determine the orientational correlation time $\xi_t$ and an ''effective'' Frank's constant $K_A$, characterizing the elastic response of topological defects to torsion in the presence of pinned particles (Fig. \ref{fig_xi}). The parameters are extracted from exponential fits $\sim e^{-t/\xi_t}$ in the isotropic fluid and algebraic fits $\sim t^{\eta_6/2}$ in the hexatic phase, where the orientational exponent $\eta_6=18k_BT/\pi K_A$ is inversely proportional to Frank's constant. In the isotropic fluid, $K_A$ is zero due to the appearance of isolated disclinations. The corresponding stress field ''absorbs'' external torsion by diffusion and/or rotation. Approaching the hexatic-isotropic~fluid transition at the temperature $\Gamma_i^{-1}\approx0.0148$, $\xi_t$ diverges, and $K_A$ jumps to the finite value $72/\pi$. In the hexatic phase, $K_A$ remains constant due to the presence of quasi-long-range orientational order: A torsion would mediate a separation of dislocations into isolated disclinations, inducing a change in the strain field at a finite stress response. Approaching the solid-hexatic transition, the elastic response to a torsion increases due to the decreasing number of isolated dislocations. Simultaneously, $K_A$ diverges. Our data indicates that in the presence of disorder, the divergent behavior of Frank's constant spreads. More precisely, $K_A$ increases at lower temperatures for higher pinning strengths which means that the hexatic-solid transition temperature $\Gamma_m^{-1}$ strongly depends on disorder, as proposed in ref.\ \cite{Nelson1983}, \cite{sachdev1984} and \cite{cha1995}. Furthermore, this implies the reduction of torsional stiffness at a fixed temperature: In the presence of pinned particles, the response to a torsional stimulus becomes more elastic.

%
%
\begin{figure}[t]
\centering
\includegraphics[width=.95\linewidth]{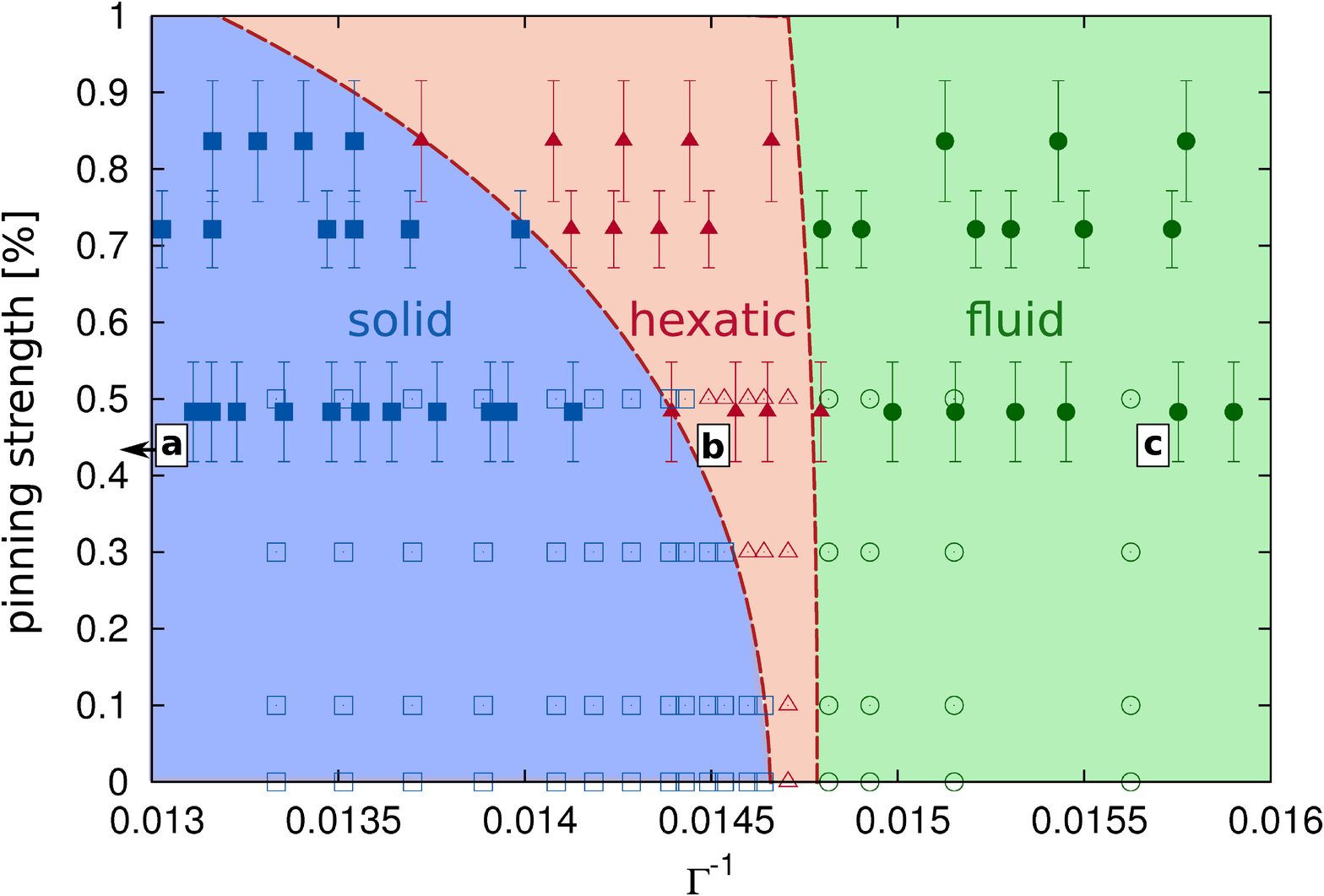}
\caption{Phase diagram indicating the solid (blue), hexatic (red) and isotropic fluid (green) phase in the parameter space of temperature $\propto \Gamma^{-1}$ and pinning strength. Full symbols represent experimental data, while open symbols correspond to simulation results. Letters indicate the location of snapshots in Fig. \ref{fig_snaps}.}
\label{fig_pd}
\end{figure}
To emphasize the consequences of these distinct characteristics at the transitions on the phase behavior of the system, the two-step melting process is mapped to the parameter plane of temperature and pinning strength. Fig.~\ref{fig_pd} shows the resulting phase diagram. In the cooling and heating cycle of the simulations no hysteresis was found, as typical for continuous transitions. The hexatic-isotropic~fluid transition is found to remain largely unaffected by pinning, the transition temperature $\Gamma_i^{-1}$ is barely shifted by disorder. In contrast, the hexatic-solid phase boundary is strongly influenced. The transition temperature is shifted significantly towards lower values for increasing numbers of pinned particles. This can be explained qualitatively considering the influence of pinned sites on the distinct symmetries: a pinned particle causes a strain field in its vicinity and therefore shifts particles to release the created stress. Orientational order can be recovered, since particles are able to adjust the orientational field $\psi_6$ to their local environment by slight displacements. However, the hexatic-solid transition is governed by a significant change in \textit{translational} order. If pinned particles are displaced from their ideal lattice position, a positional lack can only be restored by bending lattice lines. Moreover, the shear modulus is zero in the isotropic fluid \textit{and} hexatic phase which disburdens the conservation of order by adjusting the strain. As a result, the stability range of the hexatic phase widens with increasing disorder which is in accordance with theoretical predictions \cite{Nelson1983,sachdev1984}. In addition, this effect seems to become more crucial for higher disorder strengths, resulting in a curved behavior of the hexatic-solid phase boundary. This suggest the existence of a critical disorder strength, above which the system is not able to form an ordered state \cite{cha1995}, but rather becomes an amorphous solid in form of an hexatic glass \cite{kusner1994,yunker2010}, depending on the range of quenched disorder \cite{carpentier1998}.
%
%
\begin{figure}[t]
\centering
\includegraphics[width=.99\linewidth]{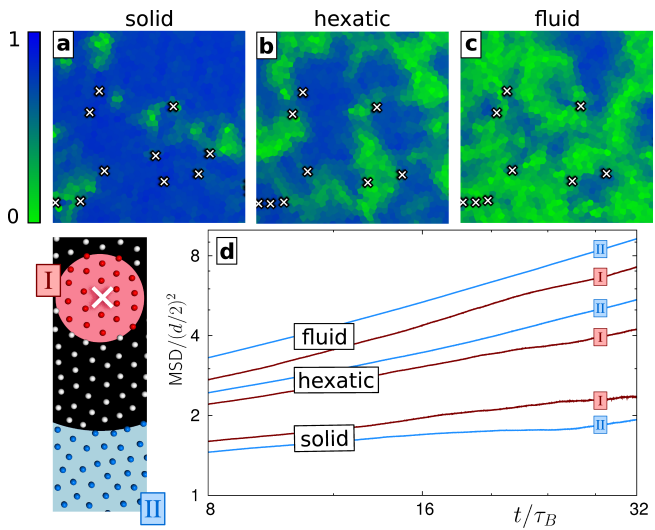}
\caption{(a-c): Snapshots of the experimental system at 0.48\% pinning, showing the local orientational order parameter $\left\langle\left|\psi_{6}\right|\right\rangle_{t}$ averaged over $\approx50\:\tau_B$ in the different phases (a: $\Gamma^{-1}=0.0117$, b: $\Gamma^{-1}=0.0143$, c: $\Gamma^{-1}=0.0154$, ). The field of view is $450 \mu\text{m} \times 450 \mu\text{m}$. Voronoi cells are color-coded according to the bar on the left. (d): Mean square displacement (MSD) calculated for particles within a distance of $8d$ around pinning sites (region I in inset) and more than $24d$ away from them (region II). Temperatures correspond to (a-c).}
\label{fig_snaps}
\end{figure}
%
%

To determine the dynamics of the orientational order in space, we illustrate the magnitude and spatial distribution of the orientational order parameter $\left\langle\left|\psi_{6}\right|\right\rangle_{t}$, averaged over a finite time window of $\approx50\:\tau_B$, see Fig.~\ref{fig_snaps},~a-c. In the solid phase, orientational order is homogeneous and persistent in time. It is only locally reduced by thermally activated, short living dislocation pairs. In the hexatic and isotropic fluid phase, the magnitudes of $\left\langle\left|\psi_{6}\right|\right\rangle_{t}$ decrease and are subject to a strongly heterogeneous spatial pattern on various length scales. This behavior can equally be observed in computer simulation snapshots, see the supplemental material. Similar heterogeneities were reported for an impurity-free two-dimensional Lennard-Jones system \cite{shiba2009}. The observed heterogeneities of the orientational order field close to $\Gamma_i$ are spatio-temporal and reflect critical-like fluctuations at the hexatic-isotropic~fluid transition, thus confirming our finding that this transition is continuous (see movie 1 and 2 in the supplemental material covering a time window two decades larger (up to $\approx 4000\tau_B$) compared to Fig.~\ref{fig_snaps},~a-c).

To exhibit the proximate effects of the pinned sites, we compare the spatial dynamics of particles in the vicinity and far away from pinning centers for an intermediate pinning strength, see Fig.~\ref{fig_snaps}~d.
While in the isotropic fluid, the mean square displacement is decreased near pinning, it is increased in the solid. The inhibited dynamics in the disordered phase can be explained by the confining character of the pinned sites. Conversely, the local dynamics in the solid is increased near pinning. This might be related to an increased probability of dislocation pair unbinding induced by quenched disorder \cite{Nelson1983,cha1995}. The crossover lies in the hexatic phase at $\Gamma^{-1}\approx0.0144$, close to the solid-hexatic phase transition which supports or finding that this transition is more affected by quenched disorder than the hexatic-isotropic~liquid one.
%
%

In conclusion, we investigated the melting transition of 2D crystals under quenched disorder in form of pinning sites. Analyzing the dynamics of the orientational correlation, we probed the disorder vs. temperature phase diagram and determined the orientational correlation time and Frank's constant. Both show divergent behavior at the corresponding phase transition, confirming the continuous melting character of the KTHNY scenario. While the hexatic-isotropic~fluid transition is rather unaffected by pinning, the transition from the solid to the hexatic phase is strongly influenced, resulting in a significant broadening of the hexatic phase. In addition, we observed spatio-temporal dynamical heterogeneities of the orientational order parameter (see movies in the supplemental material), marking critical(-like) fluctuations. In comparison to the bulk, the local dynamics of particles in the vicinity of pinned sites is decreased in the isotropic liquid but enhanced in the solid phase.
The further investigation of 2D systems with this kind of weak quenched disorder might reveal the role of critical fluctuations in the disorder-mediated melting process and also opens the field of hexatic membranes with ad-atoms/molecules. Using weak random potentials or (quasi-)crystalline structures, commensurable and incommensurable crystal transitions come into focus, and for strong disorder, crystal to amorphous solids transitions can be investigated.
%
%

We thank A. Arnold, M. Schmiedeberg and D. Hajnal for helpful discussions, as well as R. Messina and L. Assoud for providing a computer code. We thank the DFG for financial support within SFB TR6 (projects C2 and C3).
\bibliographystyle{prsty}

\begin{thebibliography}{10}
\bibitem{chi}
The magnetic susceptibility is determined via the comparison of the pair correlation function with simulations as $\chi=(5.75\pm0.05)\cdot10^{-11}\:\textrm{Am}^2/\textrm{T}$.
\end{thebibliography}

\end{document}